\documentclass[runningheads]{llncs}
\usepackage[T1]{fontenc}
\usepackage[misc]{ifsym}
\usepackage{url}
\usepackage{float}
\usepackage{amssymb}
\usepackage{graphicx,amsmath,multirow,bm,color,xr} 
\usepackage{graphicx,amstext,amsfonts,bm}
\usepackage[misc]{ifsym}

\usepackage{setspace}
\usepackage{bbding} 
\usepackage{graphicx}
\usepackage{booktabs}
\usepackage{graphicx}
\usepackage{cite}
\usepackage[colorlinks,linkcolor=red,anchorcolor=blue,citecolor=blue,CJKbookmarks=True]{hyperref} 
\usepackage{tabularx} 
\newcolumntype{L}{>{\raggedright\arraybackslash}X} 
\begin{document}
\title{Knowledge-Guided Prompt Learning for Lifespan Brain MR Image Segmentation} 
\titlerunning{Brain MRI Segmentation with Knowledge-Guided Prompt Learning} 

\author{Lin Teng \inst{1} \and 
Zihao Zhao \inst{1} \and  
Jiawei Huang \inst{1} \and  
Zehong Cao \inst{2} \and Runqi Meng \inst{1} 
\and Feng Shi \inst{2}$^{(\textrm{\Letter})}$ 
\and Dinggang Shen \inst{1,2,3}$^{(\textrm{\Letter})}$}
\authorrunning{L. Teng et al.} 

\institute{School of Biomedical Engineering \& State Key Laboratory of Advanced Medical Materials and Devices, ShanghaiTech University, Shanghai 201210, China 
\and Shanghai United Imaging Intelligence Co., Ltd. Shanghai 200230, China 
\email{feng.shi@uii-ai.com} \\ 
\and Shanghai Clinical Research and Trial Center, Shanghai 201210, China \\ 
\email{dgshen@shanghaitech.edu.cn}} 
\maketitle              
\begin{abstract} 
Automatic and accurate segmentation of brain MR images throughout the human lifespan into tissue and structure is crucial for understanding brain development and diagnosing diseases. However, challenges arise from the intricate variations in brain appearance due to rapid early brain development, aging, and disorders, compounded by the limited availability of manually-labeled datasets. 
In response, we present a two-step segmentation framework employing \textbf{K}nowledge-\textbf{G}uided \textbf{P}rompt \textbf{L}earning (KGPL) for brain MRI.  
Specifically, we first pre-train segmentation models on large-scale datasets with sub-optimal labels, followed by the incorporation of knowledge-driven embeddings learned from image-text alignment into the models. The introduction of knowledge-wise prompts captures semantic relationships between anatomical variability and biological processes, enabling models to learn structural feature embeddings across diverse age groups. 
Experimental findings demonstrate the superiority and robustness of our proposed method, particularly noticeable when employing Swin UNETR as the backbone. Our approach achieves average DSC values of 95.17\% and 94.19\% for brain tissue and structure segmentation, respectively. 
Our code is available at \href{https://github.com/TL9792/KGPL}{https://github.com/TL9792/KGPL}.

\end{abstract} 
\section{Introduction}  
Dense segmentation of brain MR images involves voxel-wise labeling from different brain tissues, \textit{i.e.}, white matter (WM), gray matter (GM), cerebrospinal fluid (CSF), into fine-grained cortical and subcortical subregions, such as 106 structures according to Desikan-Killiany (DK) atlas \cite{desikan2006automated} and Schaefer 400-parcels as described in \cite{schaefer2018local}. 
Lifespan\footnote{\textbf{Lifespan} refers to the period between birth and death, emphasizing the wide age range covered in our study.} brain segmentation is essential for quantitative brain development analysis and neurodegenerative disorders diagnosis \cite{gonzalez2016review}. Manual annotation is time-consuming and labor-intensive. Hence, an automated brain image segmentation method is urgently needed. 

Several software packages, such as FMRIB Software Library (FSL) \cite{smith2004advances} and FreeSurfer \cite{fischl2012freesurfer}, have been developed to automatically segment brain images. However, processing each subject using these tools costs several hours, and the obtained results often require further manual corrections by experienced experts. Recent advancements in deep learning, such as convolutional neural network (CNN) based U-Net \cite{milletari2016v}, Vision Transformer (ViT) based methods \cite{dosovitskiy2020image} like UNETR \cite{hatamizadeh2022unetr}, and Swin Transformer based \cite{liu2021swin} Swin UNETR \cite{hatamizadeh2021swin}, have shown promise in medical image segmentation. 

Brain morphology is dynamic in the human lifespan, leading to considerable appearance variations in brain images \cite{stiles2010basics}. For instance, CSF exhibits gradual expansion, while GM experiences progressive reduction with aging, particularly among elderly individuals with Alzheimer's disease (AD).
However, most previous efforts are established on a single time point, and primarily focus on image-based information while overlooking the complementary textual information that provides essential contextual details and insights into the pathology or anatomy, which often results in lacking generalizability over a range of ages \cite{dolz2020deep,zhang2022tw}. 
Additionally, the performance of learning-based methods is often restricted by the scarcity of annotated datasets, especially when labeled training samples for specific time points are limited in practice. To solve this problem, transfer learning \cite{pan2009survey} and fine-tuning \cite{tajbakhsh2016convolutional,roy2019quicknat} strategies have been widely adopted. 
Jia \textit{et al.} \cite{jia2022visual} proposed an efficient fine-tuning method called Visual Prompt Tuning (VPT), which incorporates a small set of random learnable parameters into the input space. 
Directly applying random parameters representing unknown information may hinder model training \cite{kang2023visual}. 

To tackle the challenges mentioned above, we present a two-step framework with knowledge-guided prompt learning (KGPL) tailored for the dense segmentation of brain MR images across diverse age groups, which are \textbf{the first} to incorporate knowledge-wise prompts into brain segmentation models. Specifically, drawing inspiration from VPT, we replace random embeddings with knowledge-driven embeddings, learned from the pre-trained text encoder in BiomedCLIP\footnote{\textbf{BiomedCLIP} indicates Biomedical Contrastive Language-Image Pre-training (CLIP) \cite{radford2021learning,zhao2023clip}, which is pre-trained on large-scale figure-caption pairs by jointly training an image encoder and a text encoder, being able to well establish the semantic connection between images and language~\cite{zhao2024chatcad+}.}, into the input space of the encoder. The knowledge-wise learnable embeddings exploit the relationship between images and brain structural and developmental-specific biomedical characteristics, aiming to empower models to learn distinguishing representations of brain data acquired from different age phases. 
Experimental results demonstrate that our proposed method outperforms state-of-the-art (SOTA) baselines, and performs robustly in different backbone networks.

\section{Methodology}  
\begin{figure} 
\includegraphics[width=\textwidth]{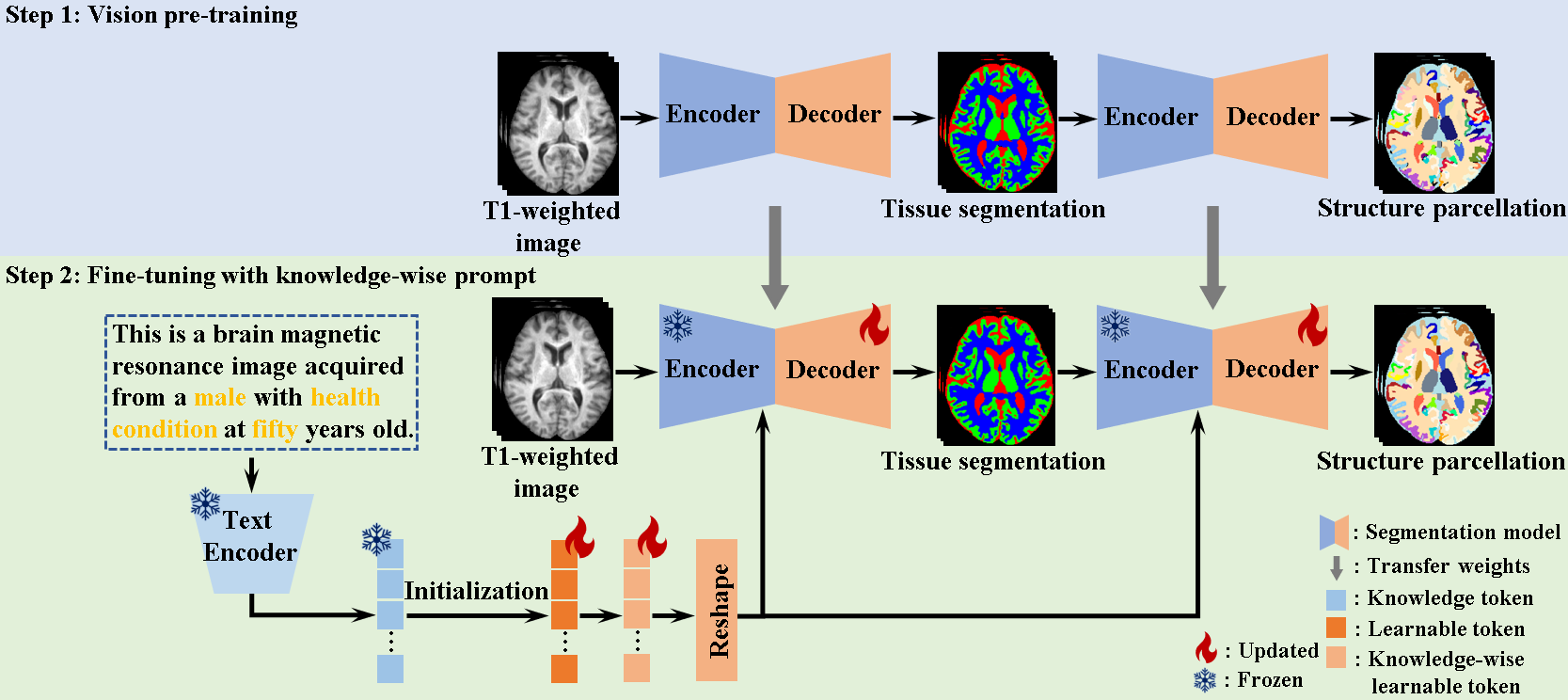} 
\caption{
Overview of our proposed two-step segmentation framework, incorporating Knowledge-Guided Prompt Learning (KGPL) for brain MRI across the lifespan. The top part shows the vision pre-training on the source domain, and the bottom part shows the refinement of the pre-trained models with knowledge-wise prompts on the target domain. 
Here, we adopt the weights learned from the source domain to initialize the model of the target domain, by only updating the learnable parameters and the decoder while freezing the encoder. 
} \label{fig1} 
\end{figure}

We aim to develop an automatic and robust approach to segment brain structure, based on brain MRI and specific biomedical attribute information. 
Fig. \ref{fig1} illustrates the overall framework. Specifically, we first pre-train segmentation models on large-scale datasets with sub-optimal annotation (Sec. \ref{sec2-1}).  
Subsequently, we refine the pre-trained models on relatively small-scale manually labeled datasets by introducing a few knowledge-driven learnable parameters into the input space, aiming to guide models to extract more crucial features of brain data spanning a large range of age. (Sec. \ref{sec2-2}).

\subsection{Vision Pre-training} \label{sec2-1} 
In this section, we utilize datasets with sub-optimal labels to establish models for segmenting brain MR images into tissue and structure, respectively. Specifically, we select one cutting-edge segmentation model as the backbone. Using the T1-weighted MR image as input, we first train the model to predict the brain tissue segmentation mask, which is supervised by the Dice loss function formulated as Eq. (\ref{func_1}). Subsequently, the predicted tissue segmentation results as input to train another model for brain structure parcellation, which is supervised by both Dice and Focal loss functions
as expressed in Eq. (\ref{func_2}). 

\begin{equation} \label{func_1} 
\text{Dice Loss} = 1 - \frac{2 \times |P \cap G|}{|P|+|G|},
\end{equation} 
where $P$ is the prediction result, and $G$ denotes the ground truth.

\begin{equation} \label{func_2}  
\text{Loss} = \text{Dice Loss} - \alpha \times (1 - p_t)^\gamma \times \log(p_t), 
\end{equation} 
where $\alpha$ and $\gamma$ are set to 100 and 0.2 based on experience.  $p_t$ is the probability of the predicted result.

\subsection{Fine-tuning with Knowledge-wise Prompt} \label{sec2-2}  
To utilize specific textual information reflecting brain properties to enhance the discriminative learning of fine-grained brain anatomical structures, we introduce a novel KGPL approach. The procedure is illustrated at the bottom of Fig. \ref{fig1}.

\subsubsection{Knowledge Prompt Generation.} \label{knowledge prompt} 
Brain morphology, such as cortical thickness, shows a strong correlation with biological attributes like brain age, sex, and disease status \cite{cole2019brain}. This textual information comes from the corresponding data acquisition centers and official websites. 
To leverage this prior knowledge for guiding feature extraction, we initially categorize subjects into age groups every decade, considering that brain morphology remains relatively similar at close ages. 
Then individual attribute details are embedded into a template to generate a descriptive sentence (\textit{e.g.}, "This is a brain magnetic resonance image acquired from a \textbf{male} with \textbf{mild cognitive impairment} at \textbf{fifty} years old"). Finally, this sentence is encoded by a pre-trained text encoder from BiomedCLIP to obtain knowledge embeddings with dimensions ($B, N, D$), where $B$ is the batch size, $N$ is identical to the number of word tokens, and $D$ is a fixed hidden dimension of 768.   

\subsubsection{Learnable Prompt Pre-initialized by Knowledge Prompt.} 
To effectively integrate prior knowledge into models, we first pre-define multiple learnable tokens (\textit{i.e.}, learnable embeddings) with zero values, matching with the dimension of knowledge embeddings. Then, the learnable embeddings are pre-initialized by knowledge embeddings through addition operation, referring to knowledge-wise learnable embeddings. 

\subsubsection{Model Fine-tuned Guided by Knowledge-wise Prompt.} 
Inspired by VPT, we serve knowledge-wise learnable embeddings as input to deep-level encoder layers, considering knowledge features are considered high-level. 
Specifically, we denote the collection of $p$ knowledge-wise learnable tokens as $P = \{ P_k \in \mathbb{R}^M || k \in \mathbb{N}, 1 \leq k \leq p\}$, where $p$ is the number of word tokens. 
To ensure that they can interact with image embeddings, we reshape them on the hidden dimension. 
For U-Net or Swin UNETR as the backbone, knowledge-wise learnable embeddings are reshaped via the Adaptive Average Pooling (AAP) operator, transitioning from ($B, N, D$) to ($B, N, 1$). Following this, they pass through a linear layer with shape ($B, C, N$), where $C$ is the number of channels; 
For UNETR, knowledge-wise learnable embeddings are first transposed from ($B, N, D$) to ($D, N, B$), then processed by a linear layer to match image embeddings' shape, except for the sequence length dimension. Next, they are inverse-transposed to restore the original dimension order. Finally, they are concatenated with image embeddings to calculate information interaction. The shape is ($B, C, (N+S)$), in which $S$ represents the dot product of the feature map's length, width, and height. 
The procedure can be formulated as:  

\begin{equation} 
[X_i, \_] = L_{i-1} ([P_{i-1},X_{i-1}]), 
\end{equation} 

\begin{equation} 
[X_{i+1}, \_] = L_i ([P_i, X_i]), 
\end{equation}  
where $X_{i} \in \mathbb{R}^{B \times C \times S}$ is image embeddings extracted by the \textit{i-1}-th encoder layer $L_{i-1}$. 
$P_i$ denotes knowledge-wise learnable embeddings at $L_i$'s input space.
During the fine-tuning step, only the image embeddings are directed to the decoder.
Notably, we initialize the model of the target domain using weights learned from the source domain, by updating only the learnable parameters and the decoder while freezing the encoder.

\section{Experiments and Results}  
\subsection{Datasets and Metrics}  
Our datasets covering various age stages are collected from multiple repositories, including CBMFM (643 subjects, aged 18-81), Alzheimer’s Disease Neuroimaging Initiative (ADNI) (1285 subjects, aged 20-97) \cite{jack2008alzheimer}, Autism Brain Imaging Data Exchange (ABIDE) (1062 subjects, aged 9-64) \cite{di2014autism}, and Adolescent Brain Cognitive Development (ABCD) dataset (458 subjects, aged 9-11) \cite{casey2018adolescent}. Detailed analysis of demographic information can be found in the supplementary material.
Most datasets processed by FreeSurfer have sub-optimal labels. To obtain accurate labels for fine-tuning and model inference, FreeSurfer is initially applied, followed by manual annotations. Each dataset is randomly partitioned into training, validation, and testing sets in an 8:1:1 ratio. 
Each subject has a T1-weighted MR image, brain tissue segmentation mask, and structure segmentation mask (106 regions). T1-weighted MR images are pre-processed by skull stripping, bias field correction \cite{tustison2010n4itk}, and intensity normalization. 
Performance evaluation utilizes two quantitative metrics: Dice Similarity Coefficient (DSC) and Average Surface Distance (ASD).

\subsection{Implementation Details} 
Experiments are conducted using MONAI \cite{cardoso2022monai} on a single NVIDIA TITAN RTX GPU with 24GB memory. Training spans 1000 epochs with early stopping, utilizing an AdamW optimizer with a learning rate of 1e-4 and regularization weight of 1e-5. We adopt a linear warmup cosine annealing learning rate scheduler. Data augmentation involves randomly flipping along three axes with a probability of 0.5. Images are cropped to (128,128,128) based on foreground regions and a resolution of 1$\times$1$\times$1 $mm^3$ using linear interpolation.

\subsection{Quantitative Comparison Analysis}  
\subsubsection{Effectiveness analysis of backbone architecture.} 
We validate the expansibility of our method based on three representative networks: (1) U-Net, a pure CNN; (2) UNETR, combining CNN and ViT; and (3) Swin UNETR, blending CNN and Swin Transformer. Their original architectures are used in our study, \textit{i.e.}, encoders have four CNN-based or twelve transformer-based blocks, decoders have four or five CNN-based blocks, and skip connections between the encoder and the decoder are included. 

\begin{table} 
\caption{The quantitative comparison results of brain tissue segmentation using different backbones, in terms of DSC and ASD (*: $p \textless$ 0.05; \Checkmark denotes to use, and \XSolidBrush denotes not to use; The best results under each backbone are in \textbf{bold}).} \label{tab1} 
\resizebox{\linewidth}{!}{
\begin{tabular}{l|l|l|l|l|l|l|l|l|l}  
\bottomrule 
\multirow{2}{*}{Backbone} & \multirow{1}{*}{Random} & \multirow{1}{*}{Knowledge} & \multirow{2}{*}{Param.} &\multicolumn{2}{c|}{CSF} & \multicolumn{2}{c|}{GM} & \multicolumn{2}{c}{WM}  \\ 
\cline{5-10}  
& prompts & prompts & & DSC (\%) & ASD (\%) & DSC (\%) & ASD (\%) & DSC (\%)  & ASD  \\  
\hline  
\multirow{3}{*}{U-Net} & \XSolidBrush & \XSolidBrush & 90.31M & 87.24$\pm$0.07* & 28.14$\pm$0.71* & 91.18$\pm$0.04* & 27.55$\pm$0.13* & 93.84$\pm$0.02* & 25.26$\pm$0.05*  \\ 
& \Checkmark & \XSolidBrush & 14.79M & 88.68$\pm$0.05* & 26.28$\pm$0.12* & 92.30$\pm$0.02* & 25.70$\pm$0.07* & 94.27$\pm$0.01* & 23.78$\pm$0.03  \\ 
& \XSolidBrush & \Checkmark & 14.79M & \textbf{89.03$\pm$0.04} & \textbf{25.59$\pm$0.11} & \textbf{93.42$\pm$0.02} & \textbf{25.02$\pm$0.07} & \textbf{95.45$\pm$0.01} & \textbf{23.27$\pm$0.03} \\ 
\hline 
\multirow{3}{*}{UNETR} & \XSolidBrush & \XSolidBrush & 92.78M & 88.24$\pm$0.07* & 28.50$\pm$0.15* & 90.09$\pm$0.05* & 28.07$\pm$0.12* & 92.20$\pm$0.03* & 30.02$\pm$0.17* \\ 
& \Checkmark & \XSolidBrush & 4.46M & 89.43$\pm$0.05* & 27.44$\pm$0.13* & 91.19$\pm$0.03* & 27.78$\pm$0.09* & 93.99$\pm$0.04* & 29.62$\pm$0.14* \\  
& \XSolidBrush & \Checkmark & 4.46M & \textbf{91.09$\pm$0.04} & \textbf{24.37$\pm$0.09} & \textbf{93.54$\pm$0.03} & \textbf{26.30$\pm$0.07} & \textbf{95.59$\pm$0.03} & \textbf{24.93$\pm$0.11}  \\ 
\hline 
\multirow{3}{*}{Swin UNETR} & \XSolidBrush & \XSolidBrush & 62.53M & 92.94$\pm$0.05* & 17.22$\pm$0.03* & 93.51$\pm$0.06* & 29.96$\pm$0.10* & 94.98$\pm$0.04* & 30.87$\pm$0.19*  \\ 
& \Checkmark & \XSolidBrush & 54.47M & 93.64$\pm$0.05* & 16.18$\pm$0.03* & 94.05$\pm$0.06* & 27.79$\pm$0.13* & 95.43$\pm$0.05* & 27.40$\pm$0.21*  \\ 
& \XSolidBrush & \Checkmark & 54.47M & \textbf{94.22$\pm$0.04} & \textbf{13.01$\pm$0.02} & \textbf{94.98$\pm$0.03} & \textbf{22.71$\pm$0.05} & \textbf{96.32$\pm$0.01} & \textbf{21.36$\pm$0.07}   \\  
\bottomrule
\end{tabular} 
} 
\end{table}

Tables \ref{tab1} and \ref{tab2} present quantitative comparison results of brain tissue and structure segmentation using three backbones, respectively. Due to page constraints, we show partial brain structure segmentation results. The complete comparison results are shown in the supplementary material.
It is evident that our approach can significantly improve the performance using different backbones, especially with Swin UNETR obtaining a high prediction accuracy compared with other backbones. 
This indicates that our method is robust in applying to various network architecture types.

\begin{table} 
\caption{The quantitative comparison results of brain structure parcellation based on different backbones, in terms of DSC and ASD (*: $p \textless$ 0.05; \Checkmark denotes to use, and \XSolidBrush denotes not to use; The best results under each backbone are in \textbf{bold}).} \label{tab2} 
\resizebox{\linewidth}{!}{
\begin{tabular}{l|l|l|l|l|l|l|l|l|l}  
\bottomrule
\multirow{2}{*}{Backbone} & \multirow{1}{*}{Random} & \multirow{1}{*}{Knowledge} & \multirow{2}{*}{Param.} & \multicolumn{2}{c|}{Left hippocampus} & \multicolumn{2}{c|}{Right hippocampus} & \multicolumn{2}{c}{Average} \\ 
\cline{5-10}  
& prompts & prompts & & DSC (\%) & ASD (\%) & DSC (\%) & ASD (\%) & DSC (\%) & ASD (\%)  \\ 
\hline  
\multirow{3}{*}{U-Net} & \XSolidBrush & \XSolidBrush & 92.31M & 93.96$\pm$0.10* & 23.51$\pm$0.47* & 92.47$\pm$0.11* & 29.69$\pm$0.40* & 88.77$\pm$0.94* & 40.41$\pm$1.52* \\ 
& \Checkmark & \XSolidBrush & 14.79M & 94.04$\pm$0.03* & 21.60$\pm$0.58* & 93.42$\pm$0.17* & 20.40$\pm$0.63* & 89.27$\pm$0.78* & 38.36$\pm$0.92*  \\ 
& \XSolidBrush & \Checkmark & 14.79M & \textbf{94.93$\pm$0.06} & \textbf{21.24$\pm$0.33} & \textbf{94.75$\pm$0.18} & \textbf{22.45$\pm$0.35} & \textbf{90.42$\pm$0.38} & \textbf{37.94$\pm$0.66} \\ 
\hline 
\multirow{3}{*}{UNETR} & \XSolidBrush & \XSolidBrush & 92.78M & 94.94$\pm$0.05* & 21.24$\pm$0.23* & 94.69$\pm$0.02* & 22.85$\pm$0.25* & 91.75$\pm$0.33* & 30.22$\pm$0.37* \\ 
& \Checkmark & \XSolidBrush & 4.46M & 93.27$\pm$0.08* & 25.48$\pm$0.31* & 93.58$\pm$0.06* & 23.15$\pm$0.36* & 91.43$\pm$0.47* & 24.75$\pm$0.63* \\ 
& \XSolidBrush & \Checkmark & 4.46M & \textbf{95.10$\pm$0.05} & \textbf{19.47$\pm$0.29} & \textbf{95.28$\pm$0.03} & \textbf{18.63$\pm$0.21} & \textbf{92.07$\pm$0.26} & \textbf{20.49$\pm$0.25} \\ 
\hline 
\multirow{3}{*}{Swin UNETR} & \XSolidBrush & \XSolidBrush & 62.53M & 96.17$\pm$0.04* & 15.10$\pm$0.33* & 96.62$\pm$0.02* & 12.71$\pm$0.22* & 93.04$\pm$0.28* & 21.28$\pm$1.86* \\ 
& \Checkmark & \XSolidBrush & 54.47M & 95.43$\pm$0.02* & 18.14$\pm$0.28* & 95.86$\pm$0.03* & 16.75$\pm$0.41* & 92.28$\pm$0.42* & 19.79$\pm$0.24* \\ 
& \XSolidBrush & \Checkmark & 54.47M & \textbf{96.65$\pm$0.01} & \textbf{11.63$\pm$0.21} & \textbf{97.57$\pm$0.01} & \textbf{9.28$\pm$0.06} & \textbf{94.19$\pm$0.21} & \textbf{14.50$\pm$0.20}  \\  
\bottomrule
\end{tabular} 
} 
\end{table} 

\subsubsection{Comparison with full fine-tuning methods.} 
Quantitative analysis of Tables \ref{tab1} and \ref{tab2} shows that KGPL using Swin UNETR as the backbone improves average DSC and ASD values by 1.36\% and 6.99\%, 1.15\% and 6.78\% for brain tissue and structure segmentation, respectively.
Statistical significance of this improvement is confirmed by a paired t-test ($p$-value $\textless$ 0.05). Notably, our method achieves optimal performance with minimal trainable parameters, particularly in UNETR, where the parameter count is reduced from 92.78M to 4.46M without sacrificing performance. 
This suggests that KGPL aids models to prioritize essential information for brain image segmentation, thereby improving efficiency and adaptability. 

\subsubsection{Comparison with random-prompts-only methods.} 
Table \ref{tab1} of brain tissue segmentation shows that our proposed KGPL using UNETR as the backbone can offer 1.87\% and 3.08\% improvements of average DSC and ASD, respectively. In Table \ref{tab2}, KGPL using Swin UNETR shows improvements of 1.71\% and 7.47\% in average DSC and ASD for right hippocampus segmentation, respectively. Importantly, our proposed method is computationally faster (3$\times$), \textit{i.e.}, extending the time from 3 days to 1 day. 
This emphasizes that knowledge-wise prompts can direct models to better comprehend and process brain image data with different characteristics and variations, thus promoting efficient convergence.

\begin{figure} 
\includegraphics[width=\textwidth] {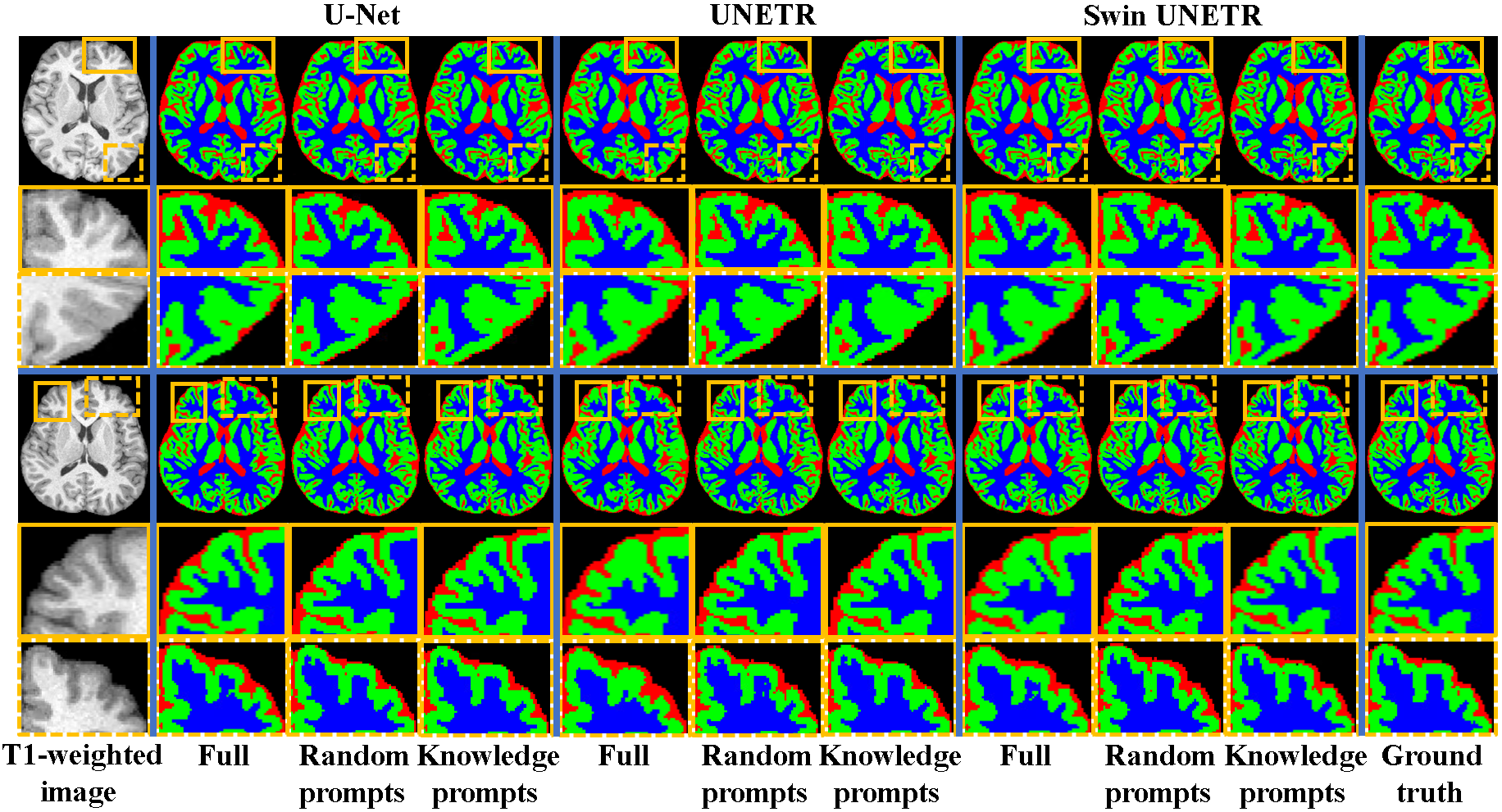} 
\centering 
\caption{Brain tissue segmentation results from different backbones are showcased in the transverse view. The first and second rows correspond to subjects aged 20 and 21, respectively. Each panel displays three segmentation results obtained by refining backbones using full, random prompts, and knowledge-wise prompts. Regions enhanced by our method are highlighted and enlarged within yellow boxes.} \label{fig2} 
\end{figure} 

\subsection{Qualitative Comparison Analysis}  
To more intuitively illustrate the superiority of our method, we display qualitative comparison results of brain tissue and structure segmentation. 

\begin{figure} 
\includegraphics[width=\textwidth]{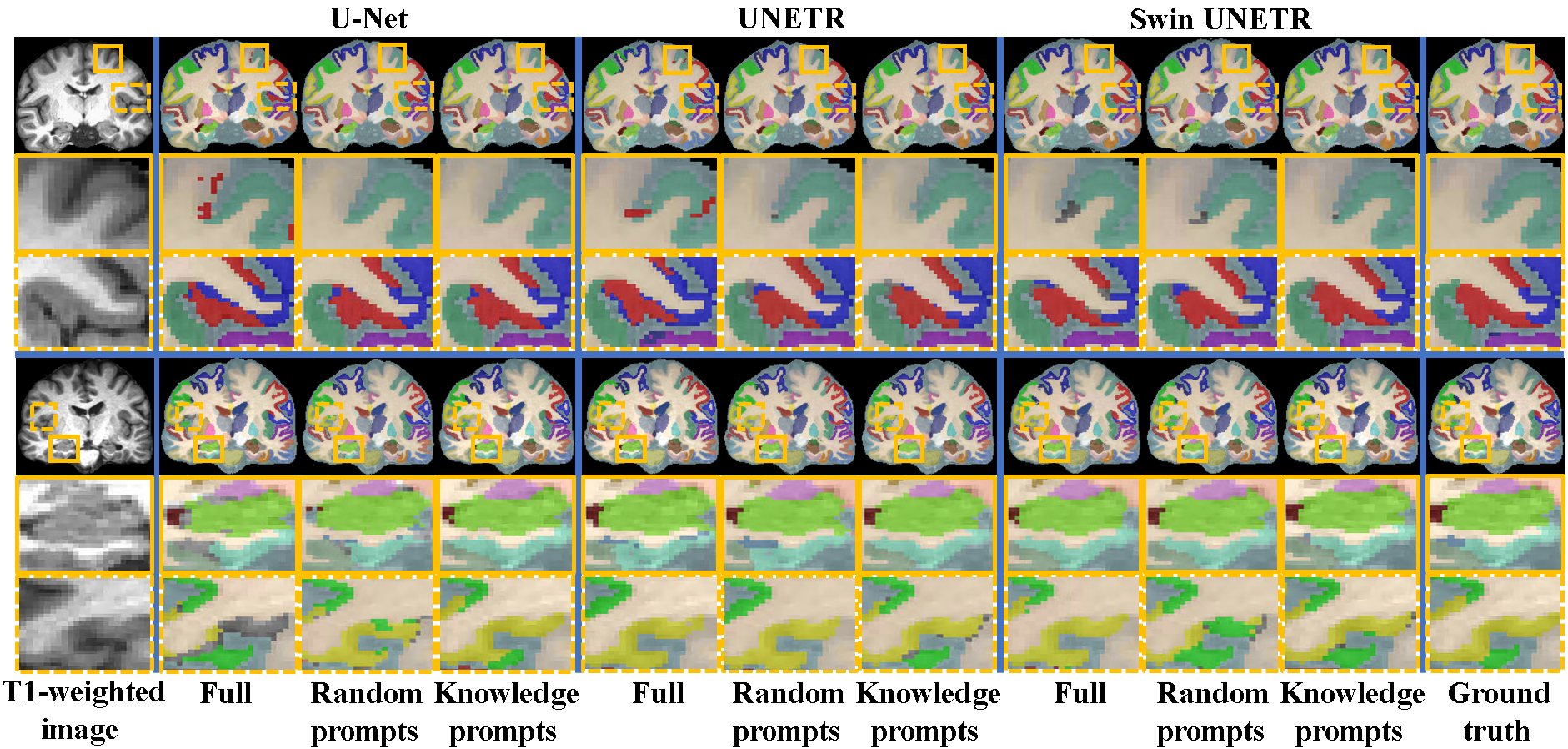}
\caption{Parcellation results of brain structure from three different backbones in the coronal view. The first and second rows correspond to subjects aged 17 and 24, respectively. Each panel displays three segmentation results achieved by refining backbones using full, random prompts, and knowledge-wise prompts. Regions improved by our proposed method are enclosed and magnified within yellow boxes.} \label{fig3} 
\end{figure} 

In Figs. \ref{fig2} and \ref{fig3}, our KGPL method yields more accurate segmentation across different backbones, with predictions closer to ground truths compared to full or random prompt-only fine-tuning.
For example, Fig. \ref{fig2} demonstrates our brain tissue segmentation closely matching the ground truth, particularly in complex brain regions like sulci and gyri. In Fig. \ref{fig3}, full fine-tuning backbones exhibit significant errors in regions like the hippocampus and cerebral cortex, while our predictions display greater precision, suggesting potential for aiding in brain disease diagnosis, such as left/right hippocampal sclerosis detection.

\section{Conclusion} 
This work presents a segmentation framework employing knowledge-guided prompt learning for brain MRI analysis throughout the lifespan. 
It stands as a pioneering effort in utilizing biomedical attribute knowledge as guidance to achieve precise segmentation for brain data. Experimental results indicate that knowledge benefits models in better learning structural variations of brain data across different age stages. 
The introduction of knowledge-wise learnable embeddings facilitates the extraction of fine-grained anatomical features and enhances training efficiency, even with small-scale datasets. 
Hence, the proposed method holds promise for clinical applications such as brain disease diagnosis. 
Moving forward, we plan to extend our methodology to infant phase datasets, enabling segmentation across the entire lifespan continuum.

\begin{credits}
\subsubsection{\ackname} This study was partly funded by National Natural Science Foundation of China (grant numbers 62131015, 62250710165, U23A20295), the STI 2030-Major Projects (No.2022ZD0209000), Shanghai Municipal Central Guided Local Science and Technology Development Fund (grant number YDZX20233100001001), Science and Technology Commission of Shanghai Municipality (STCSM) (grant number 21010502600), and The Key R\&D Program of Guangdong Province, China (grant numbers 2023B0303040001, 2021B0101420006).

\subsubsection{\discintname}
The authors have no competing interests to declare that are relevant to the content of this article. 
\end{credits}

\bibliographystyle{splncs04} 
\bibliography{Paper-1321} 

\end{document}